\documentclass[12pt]{iopart}

\usepackage{graphicx}
\usepackage{amssymb}
\usepackage{textgreek}

\begin{document}

\title[Ultra-fast prompt gamma detection in single proton counting regime]{Ultra-fast prompt gamma detection in single proton counting regime for  range monitoring in particle therapy}
\author{S.~Marcatili$^{1}$, J.~Collot$^{1}$, S.~Curtoni$^{1}$, D.~Dauvergne$^{1}$, J-Y.~Hostachy$^{1}$, C.~Koumeir$^{2,3}$, J.M.~L\'{e}tang $^{4}$, J.~Livingstone$^{1}$, V.~M\'{e}tivier$^{2}$, L.~Gallin-Martel$^{1}$, M.L.~Gallin-Martel$^{1}$, J.F.~Muraz$^{1}$, N.~Servagent$^{2}$, \'{E}.~Testa$^{5}$, M.~Yamouni$^{1}$}
\address{$^{1}$ Universit\'{e} Grenoble Alpes, CNRS, Grenoble INP, LPSC-IN2P3, 38000 Grenoble, France}
\address{$^{2}$ SUBATECH laboratory, IMT Atlantique and Nantes University, Nantes, France.}
\address{$^{2}$ GIP Arronax, Saint herblain, France.}
\address{$^{4}$ Universit\'{e} de Lyon, CREATIS, CNRS, UMR 5220, INSERM U1044, INSA-Lyon, Universit\'{e} Lyon 1, Centre L\'{e}on B\'{e}rard, 69622 Villeurbanne cedex, France.}
\address{$^{5}$ Institut de Physique Nucl\'{e}aire de Lyon, Universit\'{e} de Lyon, Universit\'{e} Lyon 1, IN2P3/CNRS, UMR 5822, 69622 Villeurbanne cedex, France.}
\ead{sara.marcatili@lpsc.in2p3.fr}
\vspace{10pt}
\begin{indented}
\item[]August 2019
\end{indented}
\begin{abstract}
In order to fully exploit the ballistic potential of particle therapy, we propose an online range monitoring concept based on high-resolution Time-Of-Flight (TOF)-resolved Prompt Gamma (PG) detection in a single proton counting regime. In a proof of principle experiment, different types of monolithic scintillating gamma detectors are read in time coincidence with a diamond-based beam hodoscope, in order to build TOF spectra of PG generated in a heterogeneous target presenting an air cavity of variable thickness.  Since the measurement was carried out at low beam currents ($<$ 1 proton/bunch) it was possible to reach excellent coincidence time resolutions, of the order of 100 ps ($\sigma$). 
Our goal is to  detect possible deviations of the proton range with respect to treatment planning within a few intense irradiation spots at the beginning of the session and then carry on the treatment at standard beam currents. 
The measurements were limited to 10 mm proton range shift. A Monte Carlo simulation study reproducing the experiment has shown that a 3 mm shift can be detected at 2$\sigma$ by a single detector of $\sim 1.4 \times 10^{-3}$ absolute detection efficiency within a single irradiation spot ($\sim$10$^{8}$ protons) and an optimised experimental set-up.
\end{abstract}

%
%
%
%
%
\section{Introduction}
Hadrontherapy makes use of light ion beams  to selectively irradiate tumours. The clear advantage of
hadrontherapy with respect to external beam radiotherapy is related to the hadron characteristic depth dose deposition profile, presenting a
maximum at the end of the range (Bragg peak); this results, in principle, in a very high ballistic precision and optimal tumour coverage.
However, uncertainties in patient tissue composition, physiological movements or transient modification of the anatomy are difficult to
assess, and significant safety margins are currently applied,
thus limiting \textit{de facto} the inherent potential of hadrontherapy (Paganetti \etal 2012). The use of an online range monitoring system is the
key to increase treatment precision and safety, and it could pave the way to the use of new irradiation fields in the vicinity of organs at risk.\\ 
Several research groups are developing range monitoring devices based on the detection of secondary particles produced by nuclear processes in the patient (Krimmer \etal 2018, Kraan \etal 2015). Our focus is on the fast detection of Prompt Gammas (PG) emitted within less than a picosecond  by nucleus
de-excitation processes  following projectile-nucleus interactions in the patient.\\ 
The principle of Prompt Gamma Timing (PGT) has been proposed by Golnik \etal 2014. It consists of measuring the  elapsed time between the proton entrance in  the patient and the detection of the PG: the proton transit time plus the photon time-of-flight (TOF) can be correlated to the PG vertex, and therefore to the proton range. In Hueso-González \etal 2015 the authors showed the potential of this technique for the detection of tissue heterogeneities and the measurements of their spatial position within the patient. In their approach using short-pulsed beams from a cyclotron, the PG TOF is measured with respect to the beam Radio Frequency (RF). This implies a series of potential limitations: the PGT spectra are typically  blurred by the time width of the accelerator bunches, by the bunch momentum spread along the beam line and possibly by any beam instability causing a phase shift in the RF signal (Petzoldt \etal 2016,  Werner \etal2019).  While the first two effects may be corrected with a non trivial calibration procedure (varying with energy), the latter is more difficult to assess and requires direct beam monitoring.\\ 
In this work we propose the use of a fast beam monitoring detector to tag in time each proton individually: this can be achieved by lowering the beam intensity to less than one proton/bunch at the very beginning of the treatment (for one or few irradiation spots),  to either confirm or interrupt the procedure in real time. In order to achieve this goal, we are developing a large area diamond-based beam-tagging hodoscope to measure the proton arrival with a time resolution better than 100 ps ($\sigma$) and its incident position with a spatial resolution of  1 mm (Gallin-Martel \etal 2018). Here we used a small size, single-channel diamond detector to show  how a fast beam trigger can result in a significant improvement of the PGT technique in terms of stability and sensitivity.\\
In the first part of this paper an experiment proving the feasibility of proton range monitoring using a PGT approach in single proton regime is reported; the second part describes a Monte-Carlo study implementing an optimised experimental set-up in order to investigate the potential of this technique in terms of range shift sensitivity.
\section{PGT experiment}
\subsection{Materials and methods}
%
\begin{figure}
\begin{center}
\includegraphics[width=0.7\textwidth]{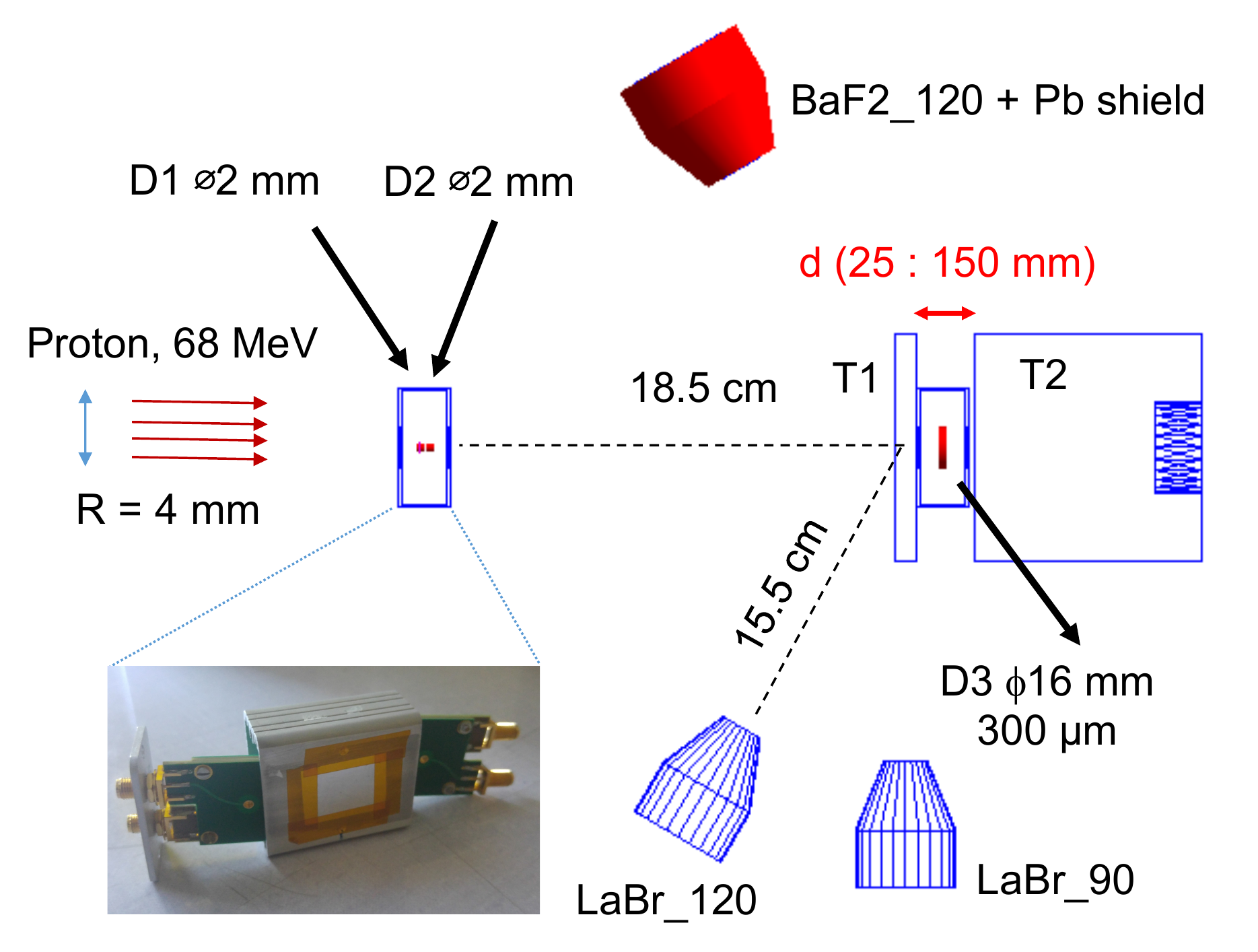}
\caption{\label{fig:setup} Schematic of experimental set-up. Target T1 is fixed in place, while T2 is progressively moved at a T1-to-T2 distance of 25, 35, 50, 70, 100 and 150 mm.  
Diamond detectors D1 (300 \textmu m thickness) and D2 (500 \textmu m thickness) are housed in the same electro-magnetic shielding box as shown in the picture inset. D3 is housed in a separate box.}
\end{center}
\end{figure}
Experiments have been performed at ARRONAX facility in Nantes.  The multi-particle isochronous cyclotron provides protons with energy up to 70 MeV for radioisotope production and R\&D activities. The working radio-frequency is 30.45 MHz. \\
Two PMMA (C$_{5}$O$_{2}$H$_{8}$) targets have been irradiated with a 4 mm radius circular shape, 68 MeV proton beam operated at an intensity of 0.37$\pm$0.03 proton/pulse: a fixed target (T1) of 10 mm thickness, and a translating target (T2) of 10 cm thickness. The experimental set-up is shown in Fig.~\ref{fig:setup}. T2 was progressively moved away from T1 (at 25, 35, 50, 70, 100 and 150 mm) in order to simulate the presence of a variable thickness air cavity and artificially increase the proton range. For all target positions the Bragg peak occurs at a penetration depth of about 1.9 cm in T2. 
The goal of this experiment was to infer the proton range shift from a TOF measurement between the (delayed) time the proton enters T1, and the time the PG reaches the detector.
 The first is measured with a diamond-based beam monitor placed upstream the targets (stop trigger) and the second by a system of gamma detectors positioned downstream (start trigger). 
\begin{figure}
\begin{center}
\includegraphics[height=5.5cm]{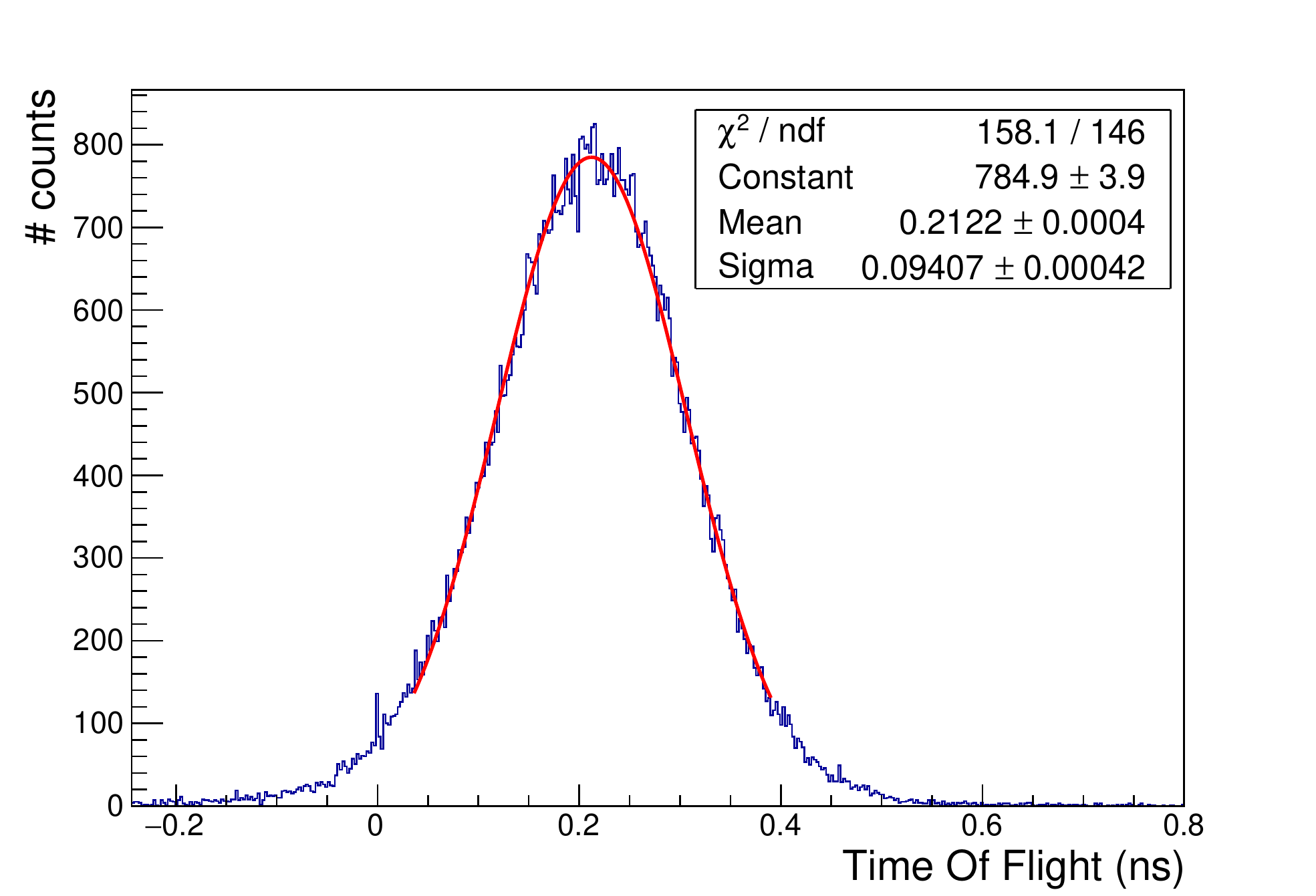}
\includegraphics[height=5.5cm]{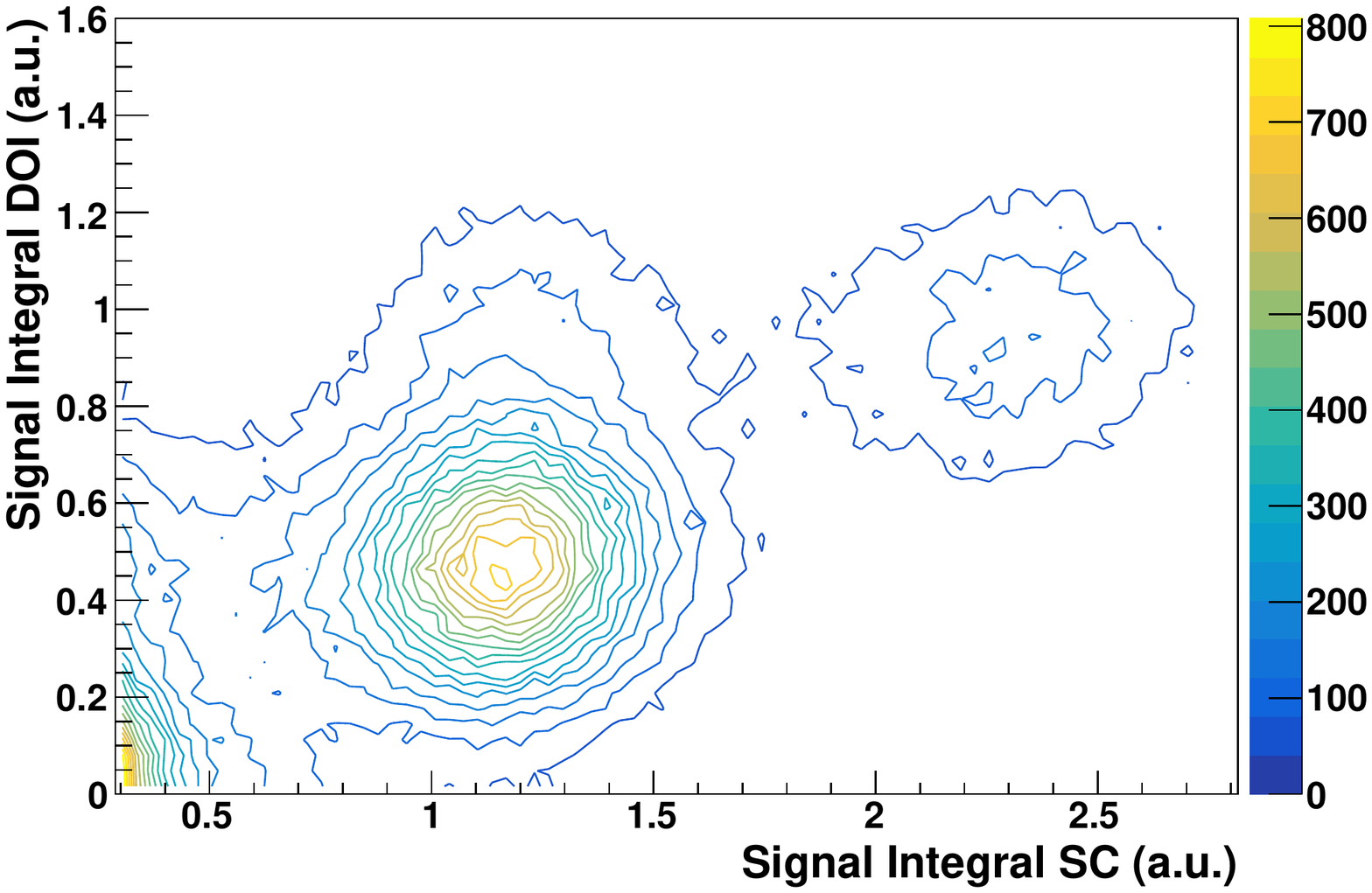}
\caption{\label{fig:tresdia} Time and energy response of diamond detectors. In the left plot, the Time of Flight between D1 and D2 for single protons is shown. Its width (94 ps rms) is interpreted as the two-diamond coincidence time resolution. The origin of time is arbitrary (different cable and connector lengths). In the right plot the energy response of D1 versus D2 is presented. The spot centered at (1.15, 0.48) corresponds to single protons traversing both diamonds. The selection of data on this region allows performing an electronic collimation of the beam. }
\end{center}
\end{figure}
\subsubsection{Beam monitoring with diamond detectors.}
Two single-channel diamond detectors have been placed in the axis of the proton beam upstream the targets: an heteroepitaxial DOI (Diamond On Iridium) from Audiatec\footnote{https://www.audiatec.de/} of $5\times 5\times 0.3$ mm$^{3}$ (D1) and a single crystal from Element Six\footnote{https://e6cvd.com/} of $4.5\times 4.5\times 0.5$ mm$^{3}$ (D2). Their effective detection area corresponds to a 3 mm diameter aluminium disk deposited on top of both sides in order to operate the diamonds in a solid state ionization chamber regime (Bergonzo \etal 2007). A bias voltage corresponding to 1 V/$\mu$m is applied to both detectors, and signals from both sides are read-out using commercial CIVIDEC C2 fast amplifiers\footnote{https://cividec.at}. 
D1 and D2  are housed in a single electromagnetic shielding box with the entrance and exit windows covered by  12 $\mu$m Mylar foils (cf. Fig.~\ref{fig:setup}). Two 1.5 mm thick FR4 Printed Circuit Boards (PCBs) hold each detector in a stack. A 2 mm diameter circular window allows clearing the detection surface on both sides of each PCB. Because of this design the diamond active area (3 mm diameter disk-shape metallisation) is larger than the open window of 2 mm diameter.
A stack of D1 and D2 allowed performing an electronic collimation of the beam and also measuring the diamond Coincidence Time Resolution (CTR). Additionally, a third diamond detector (D3) was  positioned behind T1 with the initial purpose of  implementing an electronic collimation of the proton beam, but eventually its signal was not exploited. 
\subsubsection{Prompt gamma detection.}  
Two truncated cone-shaped, Cerium-doped Lanthanum Bromide detectors (Vedia \etal 2017) of about 38 cm$^{3}$ each, have been placed downstream at 90$^{\circ}$ (LaBr\_90) and 120$^{\circ}$ (LaBr\_120) with respect to the beam axis. A Barium fluoride detector of similar shape and volume was also placed at 120$^{\circ}$ (BaF2\_120). All crystals were read-out by Photonis-XP2020 photomultipliers. 
For each T1-to-T2 distance a separate data acquisition run was performed triggering on each gamma detector. 10$^{5}$ triggers were registered for each run, with the exception of the 25~mm T1-to-T2 distance for which only 60000 triggers could be recorded with LaBr\_120.  A threshold of about 1 MeV was applied to all detectors in order to exclude 511 keV gammas from $\beta^{+}$ annihilation, and most of delayed gamma rays from nuclear reactions (Kozlovsky \etal 2002) whose TOF and emission vertex are not correlated. 
\subsubsection{Data acquisition and analysis.}
The signals from all detectors were registered for offline analysis using an 8 channels digital sampler at 3.2 Gs/s, namely the Wavecatcher (Breton \etal 2009). 
The  arrival time of protons and PGs were calculated using a digital normalised threshold of 50\% (it takes the time in the rising edge of the waveform where the amplitude reaches 50\% of the maximum), and used to build the PGT  spectrum for the three gamma detectors and each air cavity thickness. \\
In order to better interpret the experimental data, the experimental set-up was also fully implemented in a Geant4.10.5 Monte-Carlo simulation including the QGSP\_BIC\_HP\_EMY physics list. The goal was in this case to understand how the diamond packaging influenced the PGT spectra shapes. 
\subsection{Results}
\subsubsection{Diamond detector response.}
The time difference between a single proton triggering the DOI and the SC diamonds is reported in Fig.~\ref{fig:tresdia}, left. The standard deviation of this distribution is interpreted as the CTR of the two detectors: a value of 94 ps was obtained.\\
In Fig.~\ref{fig:tresdia}, right, the energy response of the DOI diamond (D1) is plotted against  that of SC (D2): the intense spot centered at (1.15, 0.48) corresponds to one-proton detected in both diamonds while the one centered at (2.31, 0.94), including about 20\% of the counts, corresponds to the two-protons signal. A small three-protons component (2\% of one-proton signal) was also detected but not included in the plot. 
The one-proton  spot presents a tail towards higher deposited energies in the DOI. This effect is associated to events for which two protons are seen by the DOI and only one by the SC. Since the detection efficiency of SC is known to be much higher than DOI's, it can be safely assumed that, for these events, one of the two protons scatters in the DOI (and/or in is packaging material) and is removed from the SC field of view. \\
With the aim of achieving the best coincidence time resolution between the proton and the PG detection, only the one-proton signal was selected for the construction of TOF spectra (\textit{single proton} selection). 
\begin{figure}
\begin{center}
\includegraphics[width=0.8\textwidth]{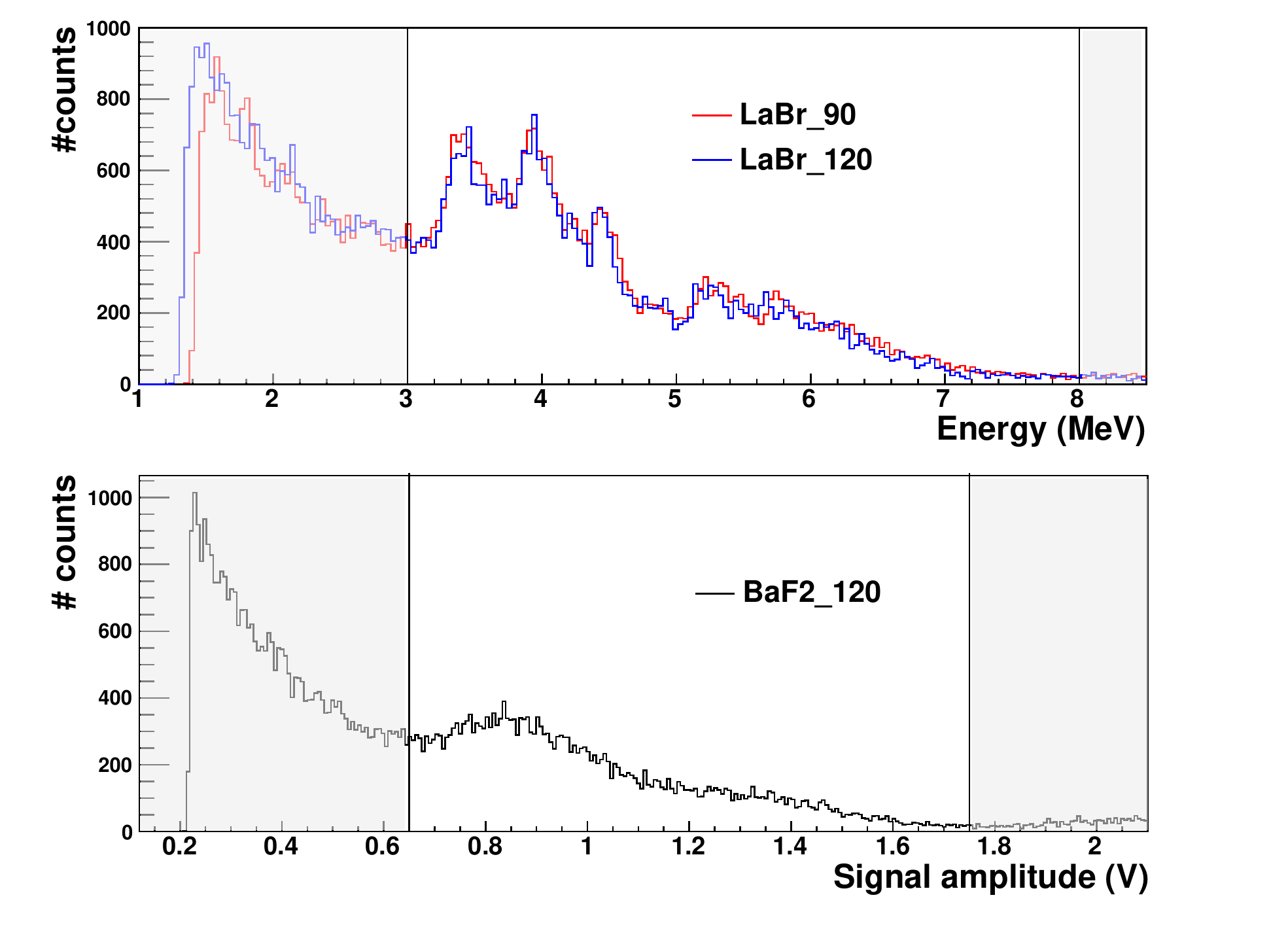}
\caption{\label{fig:pgspectra} Energy spectra obtained with the two LaBr$_{3}$ detectors, LaBr\_90 and LaBr\_120 (top), and the BaF2\_120 detector (bottom). The 3--8 MeV  energy selection window is shown in both graphs.}
\end{center}
\end{figure}
\subsubsection{PG energy spectra.}
The energy response of all gamma detectors is shown in Fig.~\ref{fig:pgspectra}. For the two LaBr$_{3}$ detectors (Fig.~\ref{fig:pgspectra}a), the signal integral was measured, then each spectrum was calibrated in energy using the three peaks recognisable in the medium energy range. The third peak is interpreted as the 4.44 MeV PG emission mainly resulting from $^{16}$O(p, x$\gamma_{4.438}$)$^{12}C$ reaction (as reported by Verburg \& Seco 2014), while the other two correspond to the single (3.93 MeV) and double (3.42 MeV) escape peaks of the 4.44 MeV gamma ray, in agreement with the limited size of our detectors. In the calibrated spectra three main regions are distinguishable. The first, with energy $<$ 3 MeV includes most of the PGs scattered in the target (Smeets \etal 2012); their TOF is not correlated to their vertex and therefore their contribution is not considered in the following analysis. The second, in the 3--8 MeV energy range,  includes the fully exploitable PG signal where, in addition to the  $\sim$4.44 MeV emissions, the lines around 6 MeV are also recognisable (a comprehensive list of PG emissions in this energy range is available in Kozlovsky \etal 2002) with their single and double escape peaks. Finally, the region with energy $>$ 8 MeV corresponds to neutrons interacting in the gamma detectors and includes the exponential tail of the PG spectrum: this background signal is cut off during analysis.\\
A different approach was implemented for BaF2\_120 since the presence of a very slow scintillation component  (630~ns)  prevented the sampling of the whole signal within the 320~ns long Wavecatcher acquisition window.  In this case the spectrum (Fig.~\ref{fig:pgspectra}, bottom) was built by measuring the amplitude of the fast scintillation component (0.8~ns). Despite the limited energy resolution of BaF2\_120 prevented us from  performing an energy calibration, the same regions present in LaBr\_90 and LaBr\_120 spectra are clearly recognisable. In this case, it was assumed that the region with 0.65 $<$ amplitude $<$ 1.75~mV roughly corresponds to the 3--8 MeV energy range.
%
\begin{figure}
\begin{center}
\includegraphics[width=0.9\textwidth]{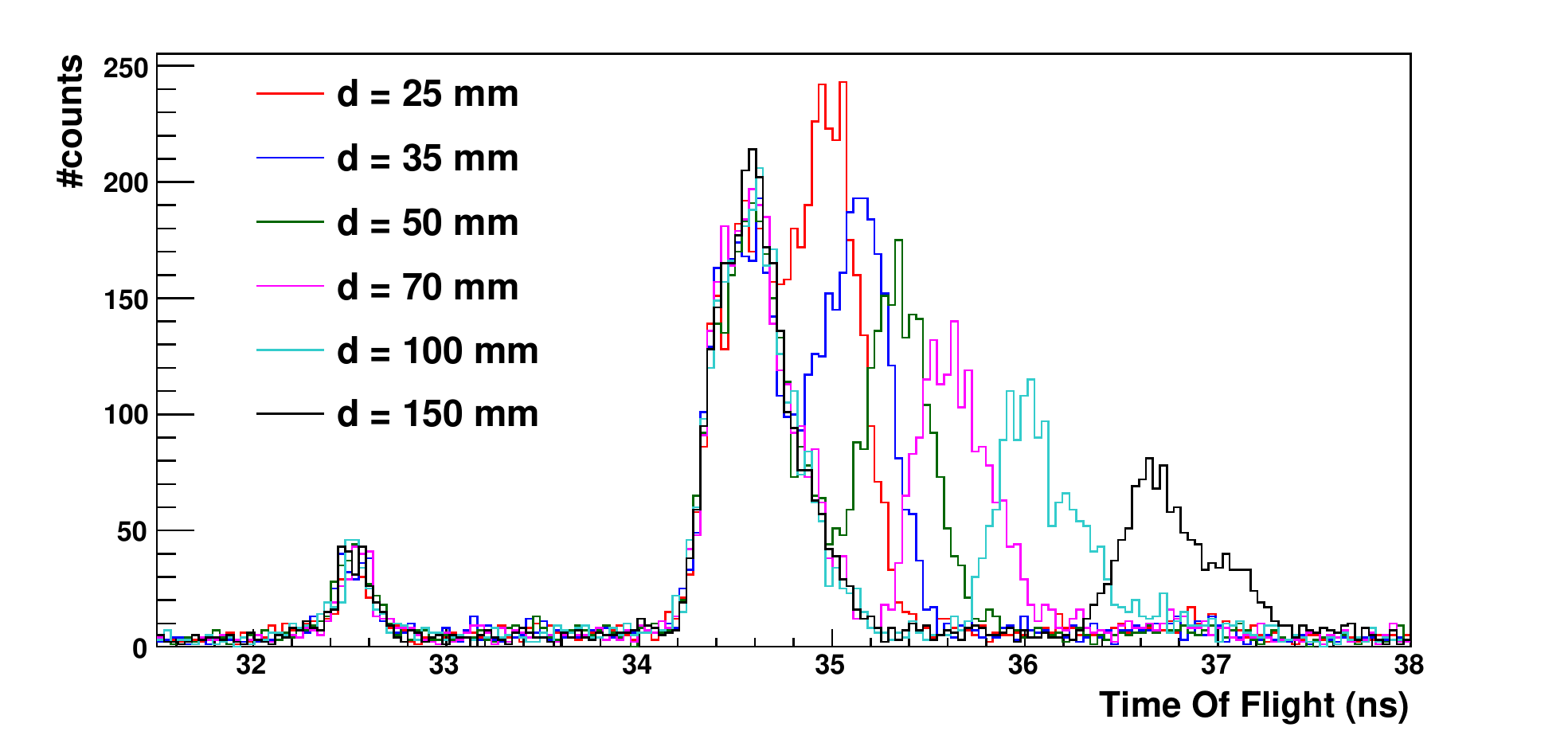}
\caption{\label{fig:movingtarget} PGT spectra obtained with the BaF2\_120 detector for different T2 positions. The absolute origin of time is arbitrary. Spectra are normalised in terms of triggers in the gamma detector (10$^{5}$): after analysis cut, approximately 5000 events remain in each spectrum. }
\end{center}
\end{figure}
\subsubsection{PG TOF spectra.}
\begin{figure}
\begin{center}
\includegraphics[width=0.5\textwidth]{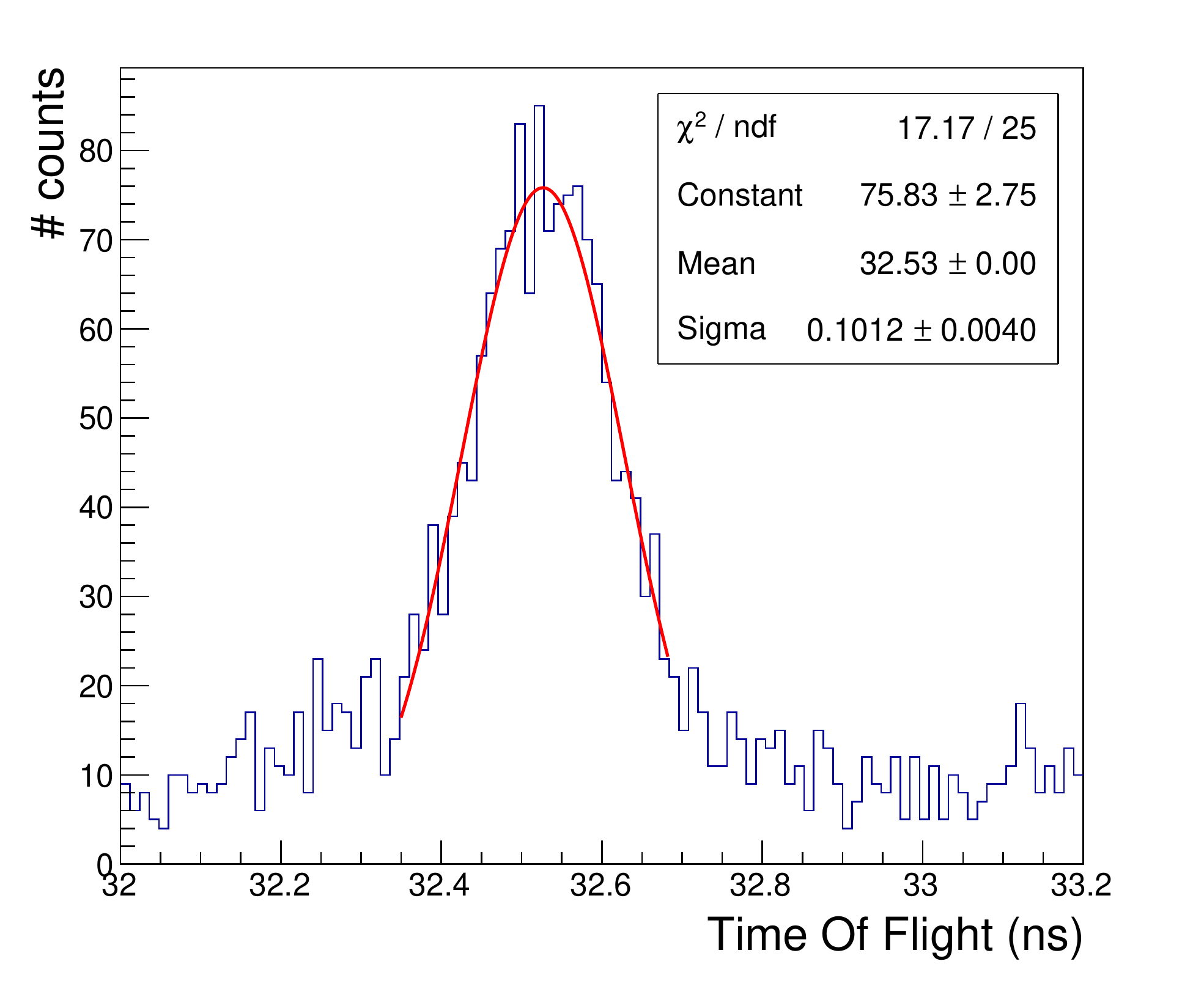}
\includegraphics[width=0.5\textwidth]{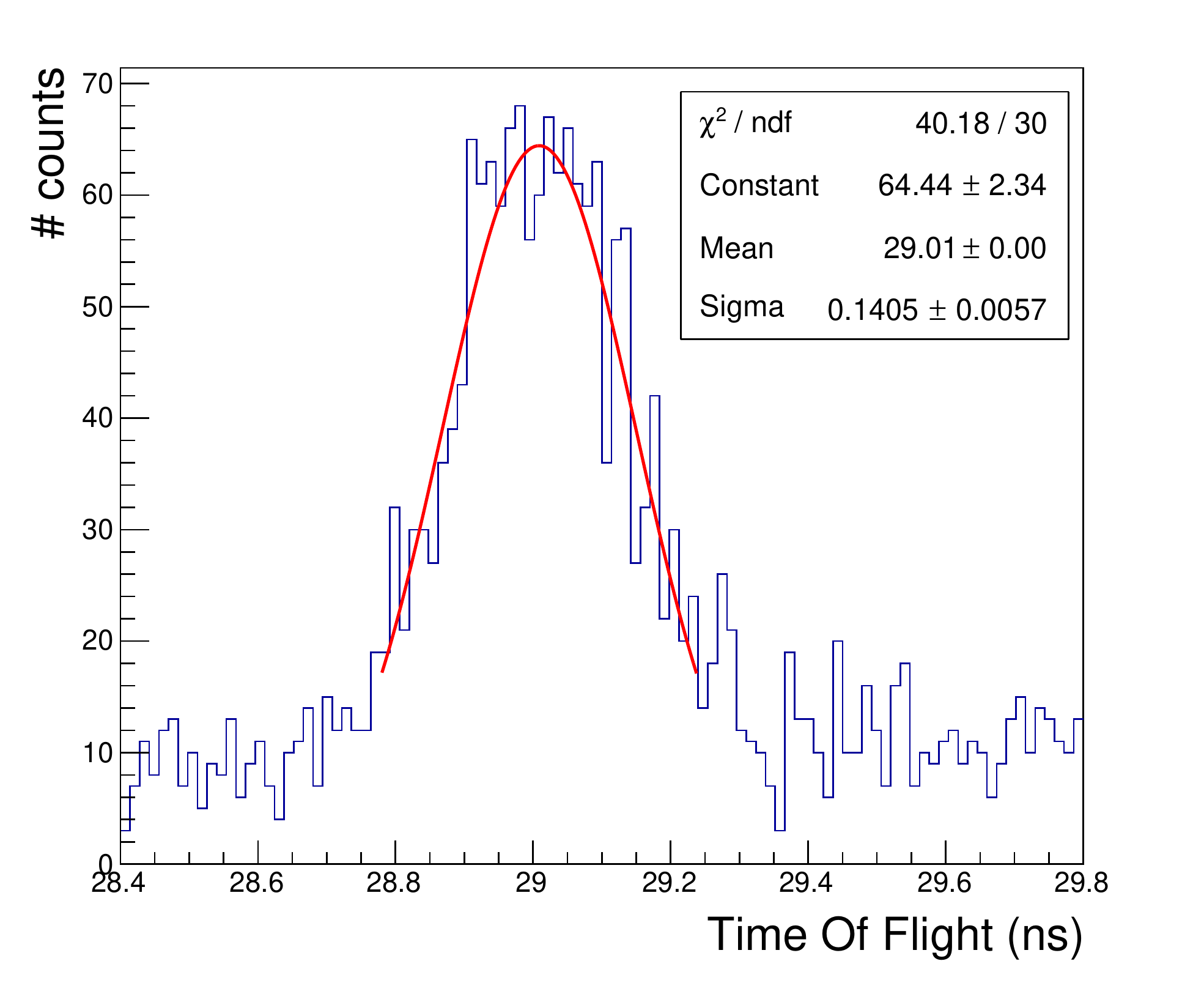}
\caption{\label{fig:tres} The gamma-proton CTR  is obtained from the PGT peak corresponding to  PG generated in the diamond detector in the case of BaF2\_120 (left) and LaBr\_120 (right). Spectra are built from 6$\times$10$^{5}$ coincidence triggers in each PG detector (acquired with a hardware threshold at E$>$1 MeV): a software energy selection in the 3-8~MeV energy range is applied on the data and only single protons hitting D1 and D2 are selected.}
\end{center}
\end{figure}

Fig.~\ref{fig:movingtarget} shows the TOF spectra obtained with BaF2\_120. The TOF values were not corrected from cable lengths, therefore the origin is arbitrary. Data are normalised in terms of total triggers in the gamma detector (10$^{5}$ in this case). The signal selection in the 3-8 MeV window allowed rejecting part of the neutron background, fast electrons originating from the targets and most of uncorrelated PGs (delayed or scattered), roughly corresponding to 30\% of acquired events.  After this energy selection and the \textit{single proton} selection are applied, each spectra approximately includes  5000 events.\\
The first peak in Fig.~\ref{fig:movingtarget} corresponds to PGs generated in the hodoscope; its dispersion is also a measurement of the system CTR. A value of 101 ps sigma was obtained for the BaF2\_120 detector  when using the single crystal diamond as stop for TOF measurement (cf. Fig.~\ref{fig:tres} left), while  CTRs of 140 ps (cf. Fig.~\ref{fig:tres} right) and 148 ps were found for LaBr\_120 and LaBr\_90 respectively. The second peak in Fig.~\ref{fig:movingtarget} corresponds to PGs generated in T1, while the third one shifts accordingly to T2 position.
\subsubsection{Measure of proton range shift.}\label{sec:distance}
\begin{figure}
\begin{center}
\includegraphics[width=0.8\textwidth]{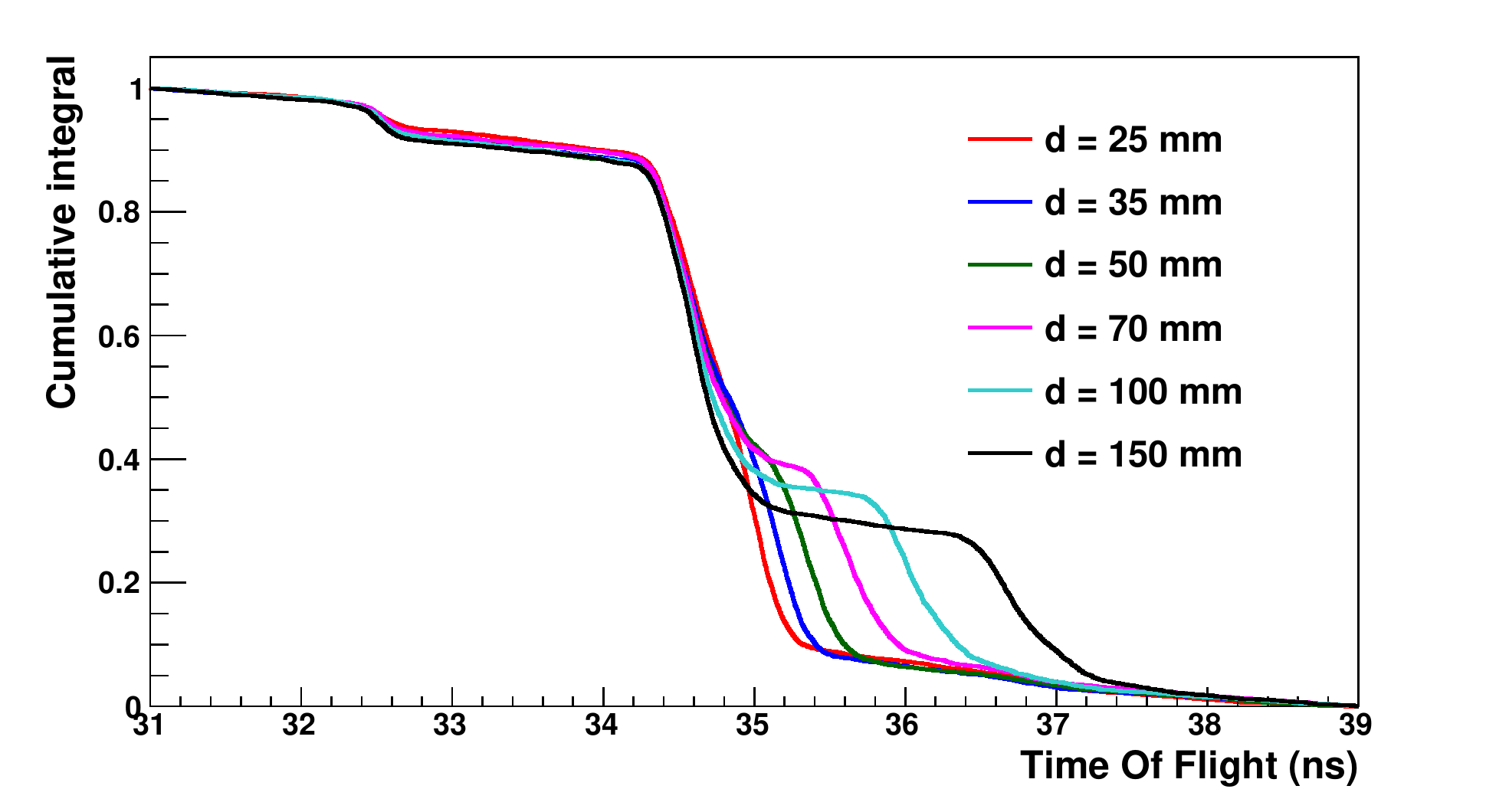}
\caption{\label{fig:integral} Cumulative integral functions obtained from a backwards integration of PGT spectra, and normalised to one.  The rapid convergence of the integral functions can be used to overcome the statistical fluctuations due to limited number of events involved in the analysis.}
\end{center}
\end{figure}

\begin{figure}
\begin{center}
\includegraphics[width=0.5\textwidth]{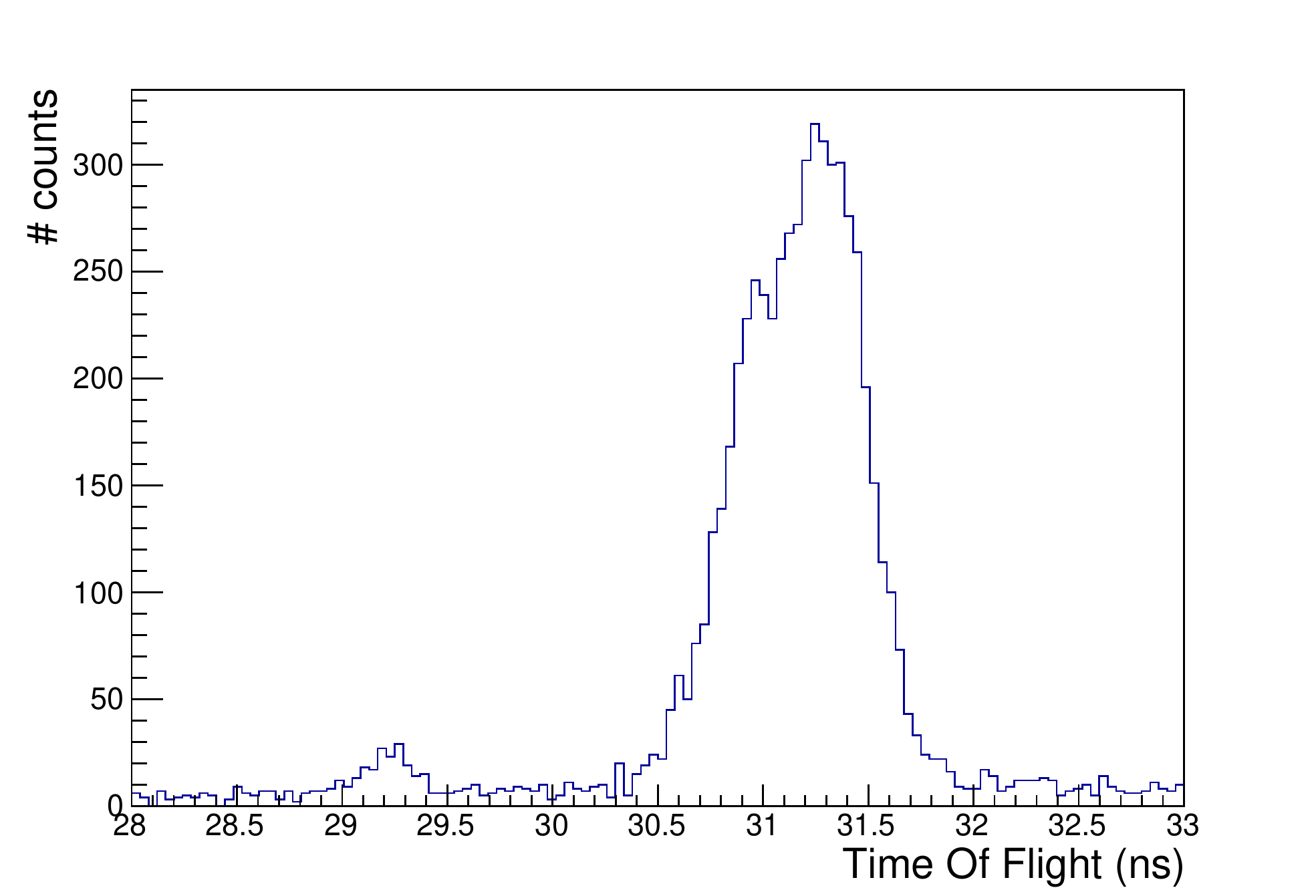}
\includegraphics[width=0.5\textwidth]{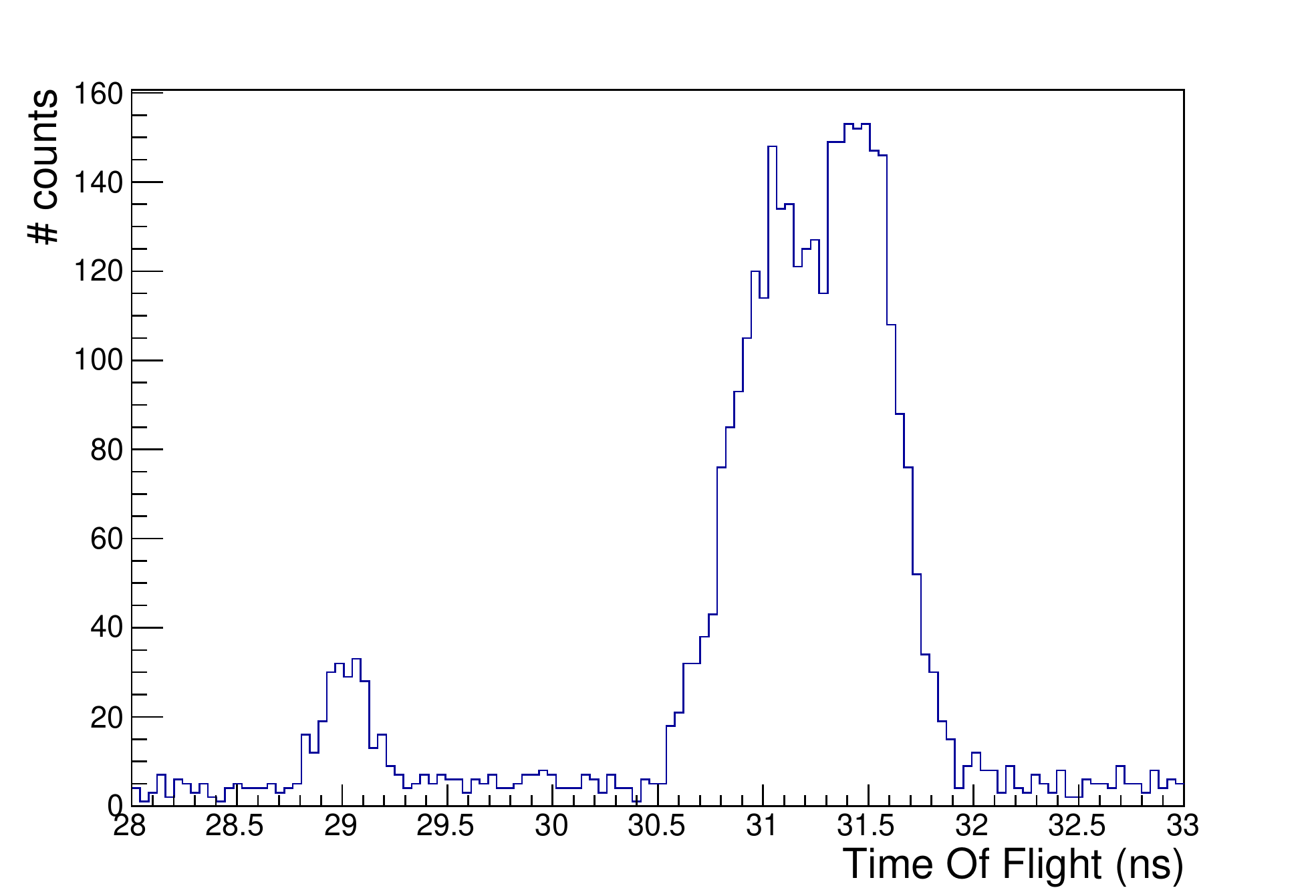}
\caption{\label{fig:120vs90} PGT spectra obtained with LaBr\_90 (left) and LaBr\_120 (right) for a T1-to-T2 distance of 25 mm. LaBr\_120 allows a better separation of T1 and T2 contributions thanks to the angular position.}
\end{center}
\end{figure}

The proton range deviation for different positions of T2 has then been measured from PGT spectra built for each gamma detector with a notion of T1-to-T2 time distance (measured in ps) robust against statistical variations of the PGT spectra, independent of PGT spectra shape, and that is defined as follows.  The PG TOF spectra are finely binned and then used to build their normalised integral functions (Fig.~\ref{fig:integral}): each peak in the original spectrum corresponds now to an inflection point. The T1-to-T2 distance is then extracted applying two arbitrary fixed thresholds (i.e. 0.8 and 0.2 for BaF2\_120 data in Fig.~\ref{fig:integral}) around the second and third inflection points in the integral functions  and measuring the relative time delay: this  corresponds to the T1-peak-to-T2-peak distance in Fig.~\ref{fig:movingtarget} and it is a measure of the air cavity thickness. 
The implementation of a numerical method to establish the threshold values has been deliberately avoided. Indeed, the PGT spectrum integral converges very quickly, making it possible to measure the T1-to-T2 distances on distributions whose shapes are only slightly affected by the limited number of events acquired. With this approach, the error in T1-to-T2 time delay is mainly systematic and depends on the PGT spectrum bin size (2 ps in this case). Conversely, the search for the exact inflection points would imply numerically deriving the integral function and therefore falling back to distributions with shapes strongly affected by statistics. This approach presents the advantage of limiting the statistical error in the measured T1-to-T2 time shift. 
\begin{figure}
\begin{center}
\includegraphics[width=0.9\textwidth]{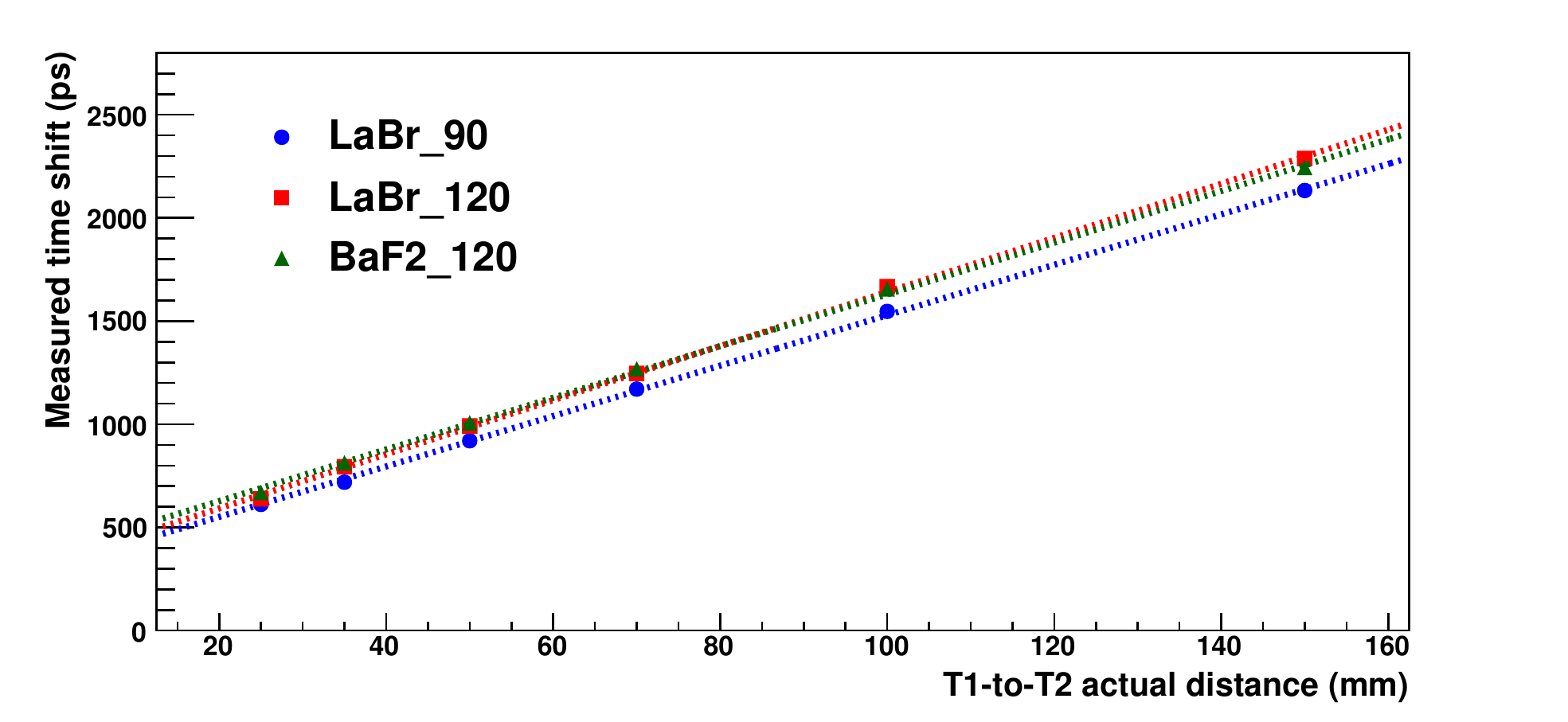}
\caption{\label{fig:distances} Measured time shifts as a function of T1 to T2 distance obtained with the three different gamma detectors. Error bars are within the point size. }
\end{center}
\end{figure}
Fig.~\ref{fig:distances} shows the correlation found between the measured and the actual T1-to-T2 distances with the three gamma detectors tested. Systematic errors are set to 1.4 ps \footnote{this corresponds to the quadratic sum of two half bins since we are computing a time difference and the time is determined by linear interpolation between two bin values} and their extension is within the point size. Detectors placed at the same angle as LaBr\_120 and BaF2\_120 have the same response since, on average, the proton-plus-PG TOF and range are the same. In this particular geometry configuration, the 120 degree position allows a better time separation of two consecutive cavity thicknesses because of the longer cumulative path travelled by the PG and the proton. This effect can also be observed by comparing, for example, the PGT spectra at 25 mm T1-to-T2 distance obtained with LaBr\_120 and LaBr\_90 for (Fig.~\ref{fig:120vs90}): in LaBr\_120 PGT spectrum, despite the lower statistics available (only 60000 triggers collected instead of 100000), the contribution of PGs from T1 and T2 is much better separated. This property governed the choice of backward angle observation of PGT by Golnik et al in their seminal paper (Golnik \etal 2014).\\
In any case, in this experiment, a range shift of 10 mm is clearly resolved with all detectors despite the higher CTRs obtained with  LaBr\_90 and LaBr\_120.  \\
\subsubsection{Monte Carlo simulations.} 
\begin{figure}
\begin{center}
\includegraphics[width=0.7\textwidth]{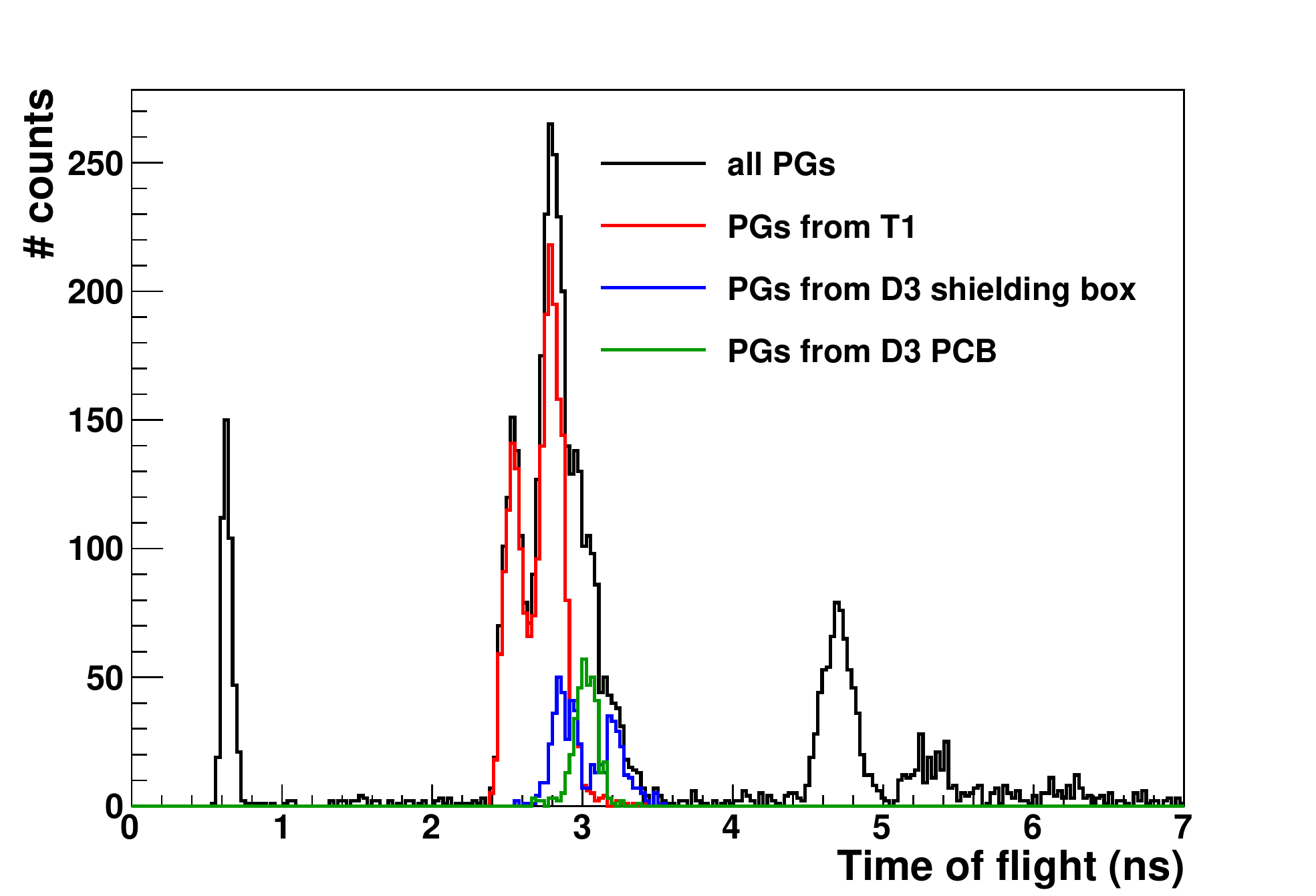}
\includegraphics[height=6.5cm]{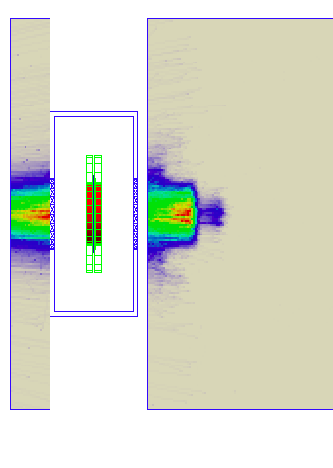}
\caption{\label{fig:mcreal} The left image shows the BaF2\_120 PGT spectrum obtained with Geant4 for a T1-to-T2 distance of 15 cm (in black). The simulation geometry replicates the actual experimental set-up. The peak at 0.6 ns corresponds to the diamond region (D1 and D2). Data between 2 and 3.5 ns correspond to the T1 target region, and those beyond 4.5 ns to the T2 region. Different components of the PGT spectrum are shown: in red PGs generated in T1, in blue and green PGs respectively generated in the shielding box and the PCB of D3. D1 and D2 packaging generates a double component in proton speed (protons going straight through diamonds versus protons traversing the packaging with higher energy loss) that results in two PG emission peaks from T1 and T2.  D3 packaging contributes to the asymmetric widening of the T1-related peak. These effects are also clearly visible in the three-dimensional representation of the simulated beam (right).}
\end{center}
\end{figure}

T1 and T2 components in the PGT spectra of Fig.~\ref{fig:movingtarget} are much broader than expected from simple TOF considerations and a system CTR of 101 ps. As 68 MeV protons travel at a speed of $\sim$10 cm/ns, the T1 component in the PGT spectrum should have a dispersion of the order of 100 ps in case of perfect time resolution, which is very far from the 1 ns obtained experimentally. In addition, T1 and T2 contributions are quite asymmetric although in principle they should not.\\
The detailed Monte-Carlo simulations of the whole experimental set-up allowed demonstrating that these characteristics are a direct effect of the diamond packaging system. 
Because of the PCB design the diamond active area (3 mm diameter disk-shape metallization) is larger than the  2 mm diameter PCB window. With a beam radius of 4 mm (measured at the entrance window of the diamond box), the proton beam is split into 3 components: i) protons crossing only the active area of diamonds, ii) protons crossing both the PCB and the active area of the diamonds, and iii) protons that are not detected by  diamonds because of their limited detection area. The latter component may generate PGs in the targets but, without a corresponding trigger in the diamond, their contribution is rejected during analysis. The first and the second components, conversely, generate a double response in the time spectra at both T1 and T2 level, due to different velocities of transmitted protons. This effect is illustrated in Fig.~\ref{fig:mcreal}, showing the simulated PGT spectrum obtained for a T1-to-T2 distance of 15 cm with the  BaF2\_120 detector, assuming a perfect time resolution (black line). In the graph, PGs generated in T1 are highlighted in red: the two peaks correspond to PGs generated by protons populations of different average speeds. Analogously, the PGT spectrum presents a double response at  T2 level (4$<$TOF$<$6 ns). At  T1 stage the shift between the two components is 260 ps, and about 560 ps at T2. 
Moreover, the additional diamond box located behind T1 (D3 cf. Fig.~\ref{fig:setup}) also generates  PGs, which results in the overall asymmetry of the PGT peak at T1. For T1 the complex time distribution is not resolved by the experimental CTR, while several components are clearly detected for T2, especially at larger T1-to-T2 distances.\\
\section{Study of proton range shift sensitivity}
\subsection{Study design}
\begin{figure}[htbp]
\begin{center}
\includegraphics[width=0.5\textwidth]{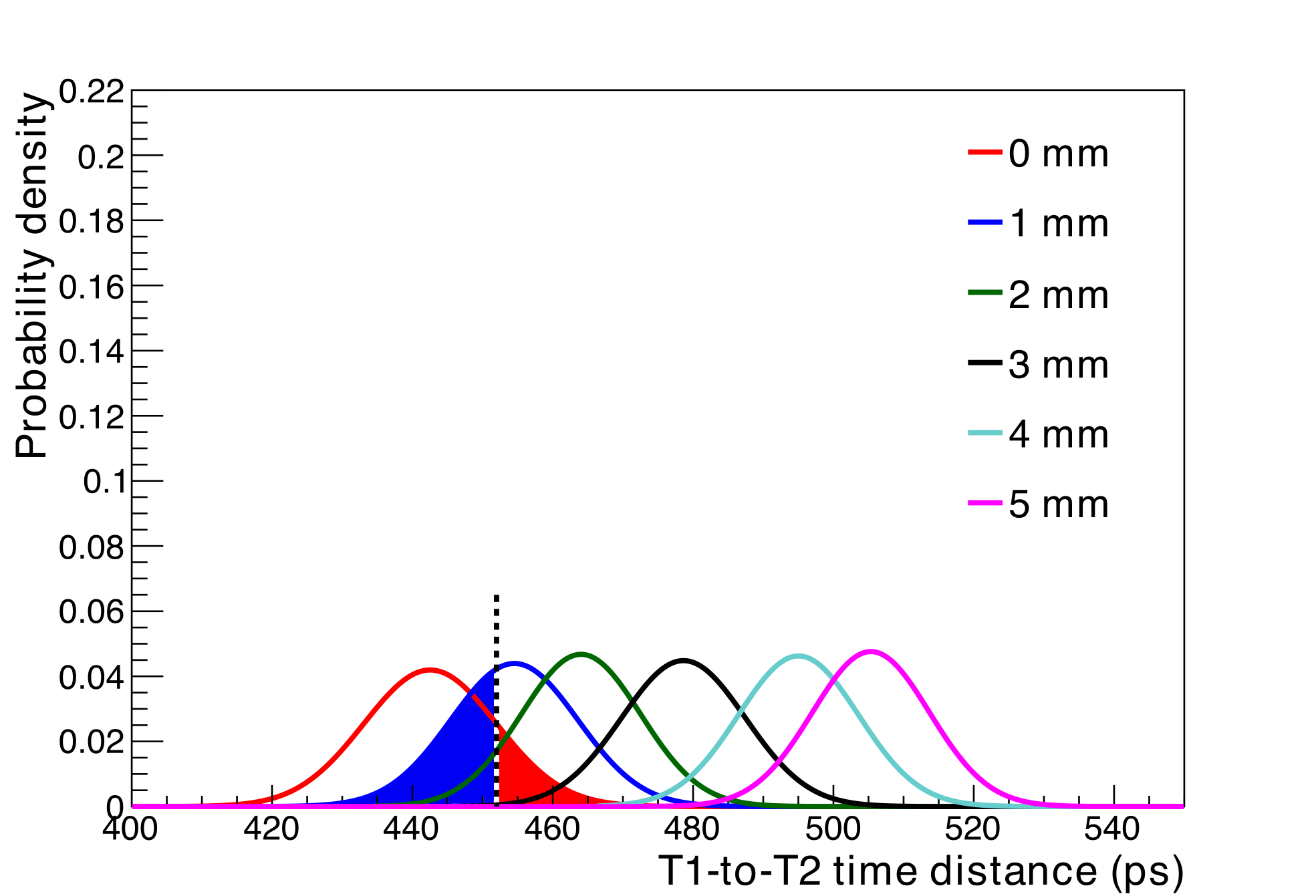}
\includegraphics[width=0.5\textwidth]{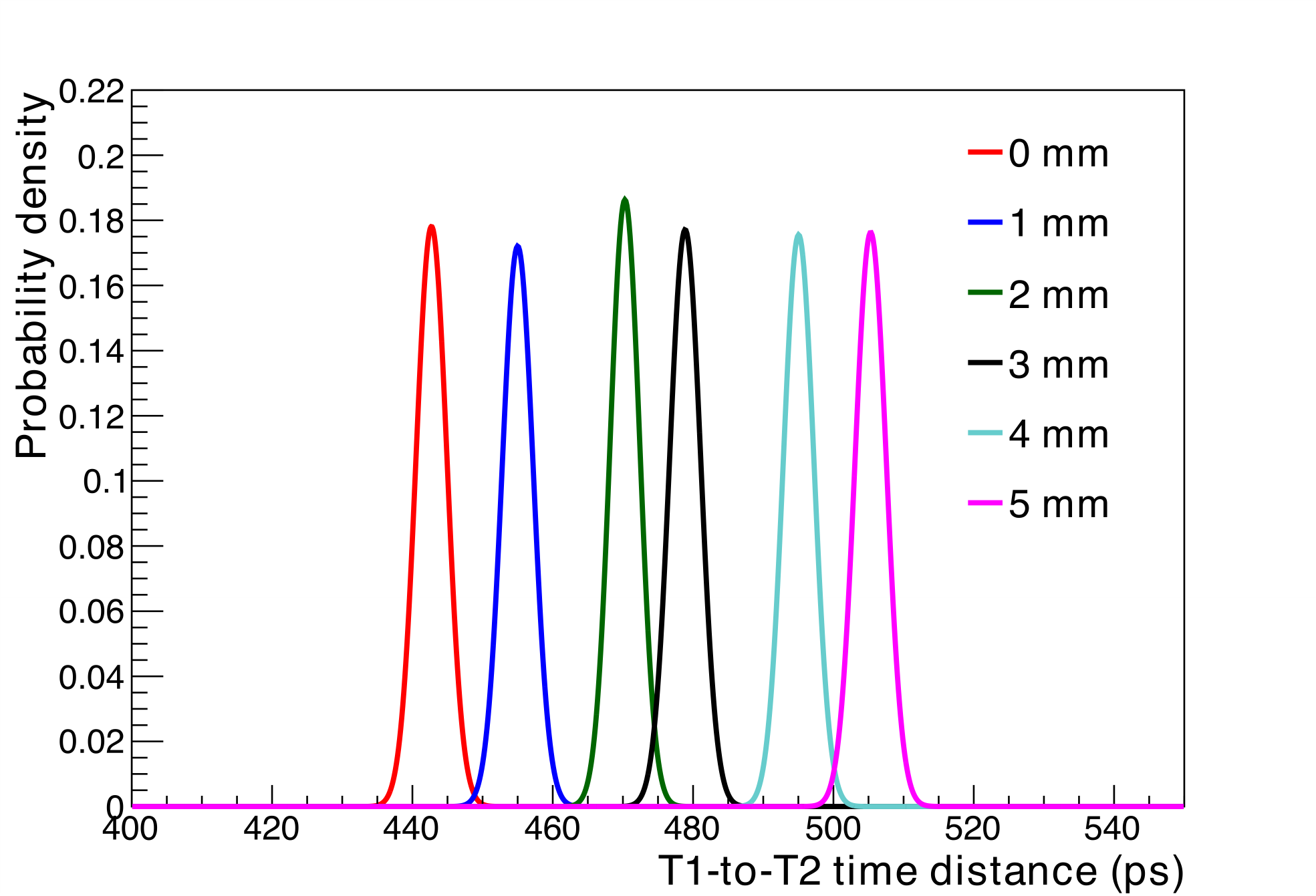}
\caption{\label{fig:pdf} 
Probability distribution functions for 0 to 5 mm shifts of T2, for 10$^{8}$ incident protons (left) and 2.6$\times$10$^{9}$ protons (right). The vertical line corresponds to a time shift value of 452.0 ps (1 $\sigma$ deviation from the mean value at 0 shift). The red area represents the type-I error, while the blue area corresponds to the type-II error for the 1 mm shift pdf.
}
\end{center}
\end{figure}

In the experiment described, proton range shifts shorter than 10 mm could not be measured because of the presence of D3 behind T1. However, the strong linearity and the limited extent of the experimental errors in Fig.~{\ref{fig:distances}}, together with the known impact of diamond packaging on PGT spectra resolution suggest that the sensitivity of our approach may be better than so far explored. 
Therefore, in order to assess the actual potential of our approach for range monitoring, additional Geant4 simulations were performed, considering an experimental set-up free of the contingent elements that degraded the experiment: the diamond (D3) behind T1 was removed, while a single large size ($4\times 4\times 0.05$ cm$^{3}$) diamond detector acting  as hodoscope, as the one under development in our collaboration (Gallin Martel \etal 2018), was included. The hodoscope was set closer to T1 (5 cm distance) as advisable in clinical conditions in order to minimize the beam angular divergence introduced by proton interactions in the diamond. A 1.8 mm radius-circular beam of 68 $\pm$ 0.1 MeV was modelled. Only the two LaBr$_{3}$ detectors have been included in this simulation. The T2 target, initially positioned at 25 mm distance from T1, has been displaced at 26, 27, 28, 29 and 30 mm distance to study the system sensitivity to small range shifts. A total of 2.6$\times 10^{9}$ primary protons was simulated; data were analysed considering randomly selected event subsamples of progressively smaller size (down to 10$^{8}$ protons, roughly corresponding to a single intense distal spot) in order to investigate the sensitivity dependence on the primary protons number. 
Realistic PGT spectra corresponding to LaBr\_90 and LaBr\_120 were generated from simulated data by
smearing the simulated TOF (between the gamma detector and the hodoscope) with a  gaussian distribution of 100 ps standard deviation, on an event-by-event basis. This convolution has a major impact on PGT spectra shapes: convolving the same subsample twice using two different random seeds produces two statistically independent distributions. For each spectrum only PGs with energies in the 3-8 MeV window were considered, in line with the analysis carried out on the experimental data.  \\
For each subsample, 10000 toy experiments were generated to build an equal number of possible experimental PGT spectra. Each spectrum was integrated and the T1-to-T2 time shift was estimated applying a set of fixed thresholds consistently with the data analysis carried out on the experimental data and described in section \ref{sec:distance}. The T1-to-T2 time shift obtained were then used to build the probability distribution function (pdf) of the  T1-to-T2 time shift, for each simulated T2 shift  (from 1 to 5 mm considering a T1-to-T2 distance of 25 mm as the null shift) and different levels of statistics. Fig~\ref{fig:pdf} shows the pdfs obtained with LaBr\_30 in the case of $10^{8}$ (left) and 2.6$\times 10^{9}$ incident protons. 
Fixing a type-I error ($\alpha$) at 1, 2 and 3 $\sigma$ on the pdf corresponding to the null shift (T1-to-T2 distance of 25 mm), the pdfs for different T2 shifts can be used to calculate the corresponding type-II errors ($\beta$), that is the probability that we fail to reject a null shift when the proton range is actually shifted. In Fig.~\ref{fig:pdf} left, areas corresponding to type-I (red) and type-II (blue) errors at 1$\sigma$ are shown for reference in the case of a T1-to-T2 distance of 26 mm (1 mm range shift).
For a given level of statistical significance it is  therefore possible to build curves describing the variation of this probability as a function of the number of incident  protons.
\subsection{Results}
\begin{figure}[htbp]
\begin{center}
\includegraphics[width=0.5\textwidth]{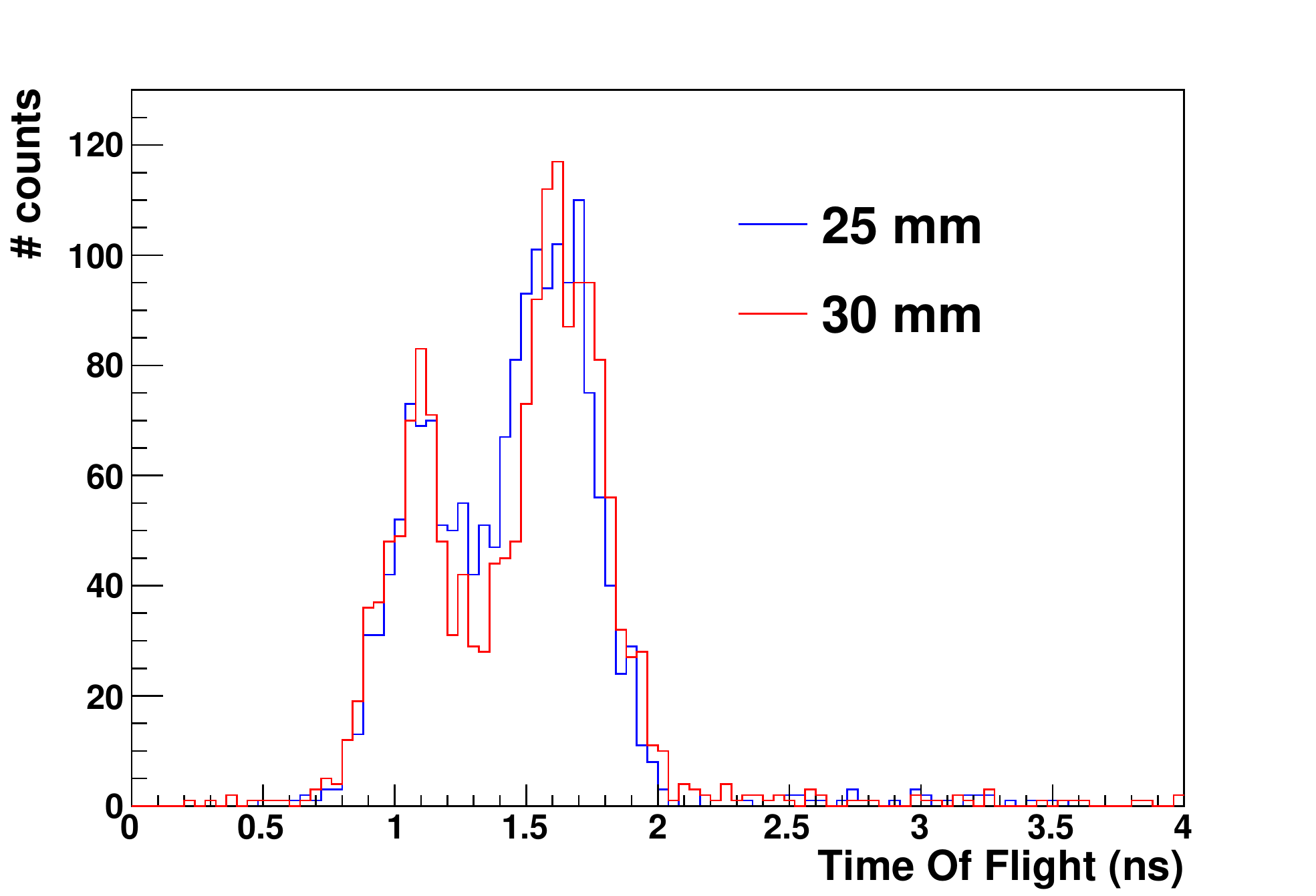}
\includegraphics[width=0.5\textwidth]{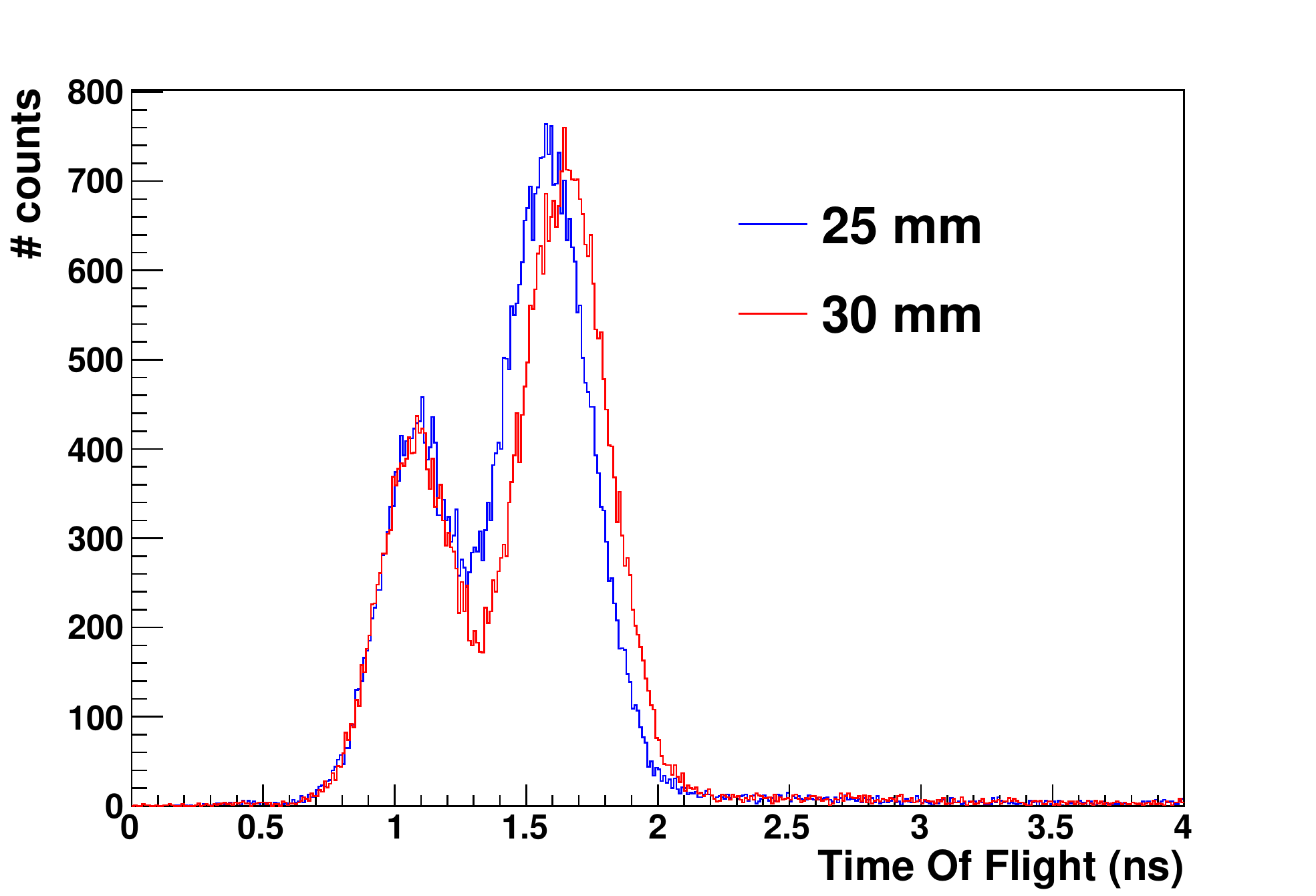}
\caption{\label{fig:simuidealpgt} Simulated PGT spectra obtained with LaBr\_90 under ideal conditions (the diamond hodoscope has no packaging and D3 is removed) and realistic time resolution: on the left, PGT spectra are obtained with 10$^{8}$ incident protons, while spectra on the right plot correspond to 2.6$\times$10$^{9}$ incident protons. }
\end{center}
\end{figure}

Typical simulated PGT spectra obtained with LaBr\_120 are shown in Fig.~\ref{fig:simuidealpgt} in the case of two T1-to-T2 distances and for both the reference statistics of $10^{8}$ incident protons (left) and a larger high statistics of 2.6$\times10^{8}$ protons (right). Once the diamond packaging is removed, the T1 and T2 components in the PGT spectra present are much narrower (c.f. PGT spectra in Fig.~\ref{fig:movingtarget}), and their width is in agreement with proton kinematics: the FWHM of T1 component is in this case of $\sim$150 ps.\\
\begin{figure}[htbp]
\begin{center}
\includegraphics[width=0.9\textwidth]{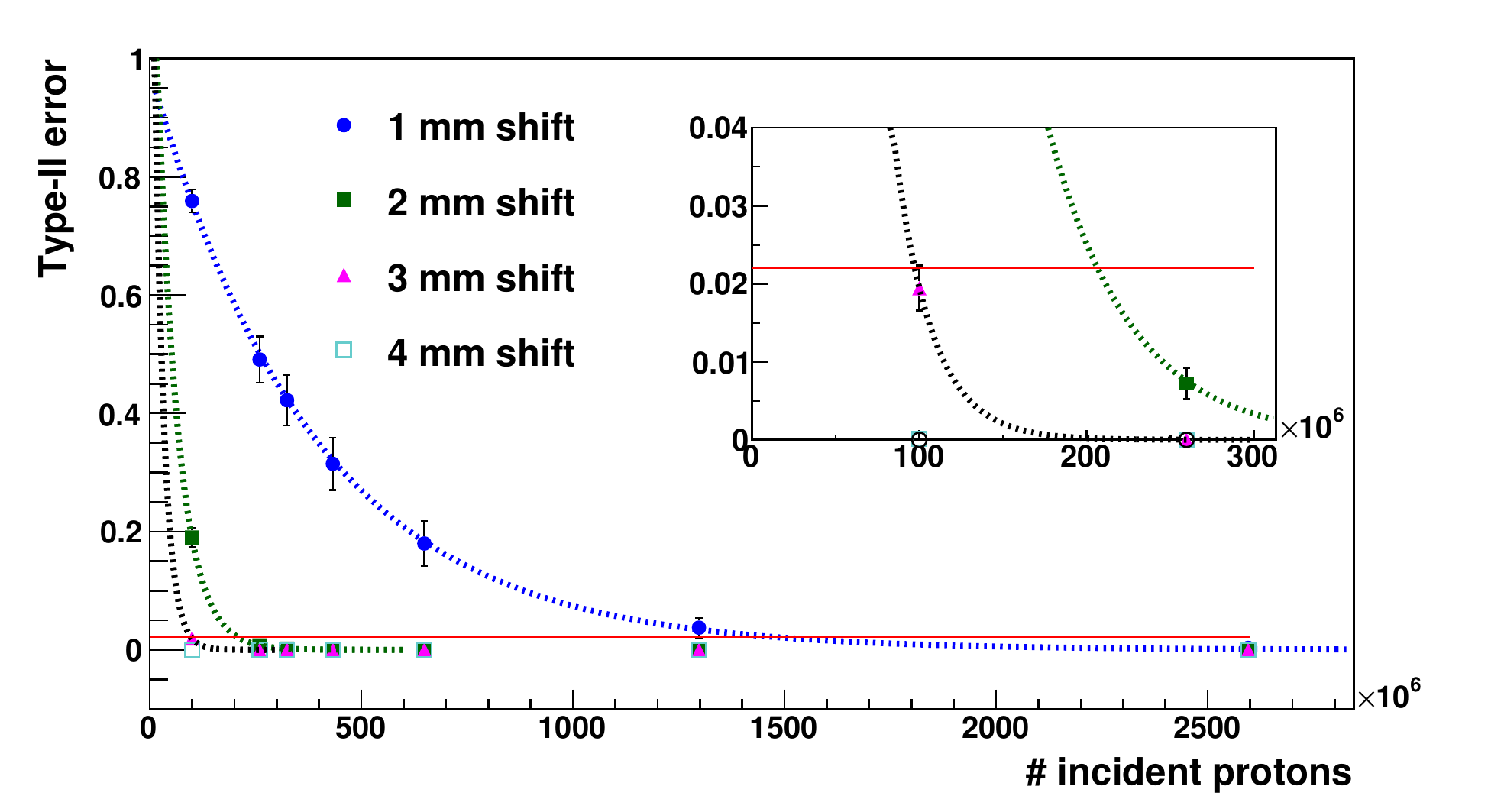}
\caption{\label{fig:stat} Type-II error at 2$\sigma$ as function of the number of primary protons used for range measurement with LaBr\_90. The type-II error indicates the probability that we fail to reject a null shift when a shift  is actually present. The red horizontal line represents the 2$\sigma$ value ($\beta$=0.023), for reference. The dashed lines are displayed as eye-guides to ease data reading. }
\end{center}
\end{figure}
Ten thousands spectra equivalent to those shown in Fig.~\ref{fig:simuidealpgt} were used to build the pdfs for the different T1-to-T2 time shift and to carry out the range sensitivity study. 
In Fig.~\ref{fig:stat} the type-II error for $\alpha = 2 \sigma$ (unlike in Fig.~\ref{fig:pdf}, where 1 $\sigma$ was chosen for better visibility), obtained for LaBr\_90 is plotted against the number of primary protons included in the analysis. The red horizontal line represents the 2$\sigma$ value ($\beta$=0.023), plotted for reference, while the dashed lines are only displayed as eye-guides to ease data reading. From Fig.~\ref{fig:stat} it follows, for example, that a 3 mm shift can possibly be detected with $10^{8}$ primary protons at 2 sigma. This results corresponds to a PGT spectrum of 1850 entries after a PG energy selection between 3 and 8 MeV is applied. Error bars represent the systematic error associated to the integration of pdfs with finite binning. Similarly, type-II errors have been computed for different levels of statistical significance and both LaBr detectors. Results are listed in table \ref{tab:table4}. From these data LaBr\_90 seems to perform better in terms of sensitivity than LaBr\_120 contrary to what was discussed in section \ref{sec:distance}. Nevertheless, some considerations should be made to better contextualise the results obtained via Monte-Carlo simulation. First, all type-II errors obtained for LaBr\_90 and LaBr\_120 in this analysis are very close. For example, considering a 2$\sigma$ level of significance and a statistics of $10^{8}$ incident protons, the type-II errors at 3 mm shift for LaBr\_90 and LaBr\_120 are 0.019$\pm$0.008  and 0.026$\pm$0.010 respectively. Both results could be interpreted as being either lower or higher than the 0.023 probability corresponding to 2$\sigma$, but for simplicity's sake we chose to classify type-II errors according to their average values. In any case, this kind of analysis ultimately depends on patient/phantom geometry and should  only be trusted as indicative. 
We have an example in this work of how an heterogeneous geometry may affect the sensitivity assessment. In the high statistic pdfs obtained for LaBr\_120 (Fig.~\ref{fig:pdf} right), all mean distribution values should be, in principle, evenly spaced since the distributions are built by increasing the air cavity by the same amount at each step (1 mm). The observed deviation from the ideal response is related to variations in materials compositions and thickness in the PG paths from their vertex to the detector. Indeed,  Fig.~\ref{fig:setup} shows how the LaBr\_120 solid angle covers a region in which phantom characteristics strongly vary as a function of T1-to-T2 distance. 
This is not the case for LaBr\_90 for which, being directed towards the center of T1, phantom characteristics only slightly vary; as a result, pdfs obtained with LaBr\_90 (not shown here) are evenly spaced in the T1-to-T2 distance axis. The net, and purely accidental, result is that in our analysis the 3 mm pdfs is closer to the 2 mm one for LaBr\_120, directly translating into a lower estimated sensitivity for this detector. \\
\Table{\label{tab:table4} Range shift sensibility expressed as type-II error at 1, 2 and 3$\sigma$, calculated for LaBr\_90 and LaBr\_120 and two different numbers of primary protons. 10$^{8}$ protons roughly correspond to a single intense (distal) irradiation spot. According to simulations, 10$^{8}$ primary protons generate 1850 and 2250 events in the PGT spectra for LaBr\_120 and LaBr\_90 respectively, after the 3-8 MeV energy selection is applied. }
\br 
		&   \multicolumn{3}{c}{Range sensitivity (mm)}& \multicolumn{3}{c}{Range sensitivity (mm)}\\
		&   \multicolumn{3}{c}{ with $10^{8}$ protons}& \multicolumn{3}{c}{with $2.6\times10^{8}$ protons}\\
Detector& 1$\sigma$&	2$\sigma$&	3$\sigma$&	1$\sigma$&	2$\sigma$&	3$\sigma$\\
\mr
LaBr\_90& 1& 3& 5& 1& 2& 3\\
LaBr\_120& 2& 4& 5& 1& 3& 4\\
\br 
\endTable 
\section{Discussion}
%
The main element of novelty in our approach consists in the use of a fast beam hodoscope for single proton counting.
On the one hand, the presence of a beam hodoscope is the key to avoid  continuous (for each energy) beam calibration procedures that are laborious and still cannot prevent an unexpected beam instability to occur during treatment.
On the other hand, its use in single proton regime is essential if CTRs of 100~ps or less are targeted, as it allows getting rid of the time jitter introduced by the finite bunch time width (a few ns for cyclotrons, tens of ns for synchrotrons). 
As a consequence, the proposed technique implies a reduction of the beam intensity ($\sim$ 1 proton/bunch) during a selection of one or few intense beam spots, to possibly interrupt the treatment at its very beginning in case a disagreement with the treatment plan is detected. Intensity reduction could be achieved by mechanically inserting  pepperpot-type grids in the beamline. 
Considering standard cyclotron accelerators providing proton bunches every 10 ns or so, about 10 s would be needed to deliver 10$^{8}$  protons at an intensity of 1 proton/bunch. 
Thus the time necessary to change the beam intensity represents a few seconds, and the additional time required for each treatment should not significantly impact the clinical workflow (Pausch \etal 2019). \\ 
Other than allowing a real time monitoring of the treatment, the use of ultra fast PG detectors also consents relaxing the detector requirements in terms of energy resolution. This leads to remarkable advantages such as a reduction in detector complexity and cost. 
We have shown in this work that a very limited energy resolution is sufficient to guarantee a good signal to noise ratio. In high time resolution PGT the main sources of background are neutrons and PG scattered in the target, whose vertex and TOF do not correlate: the firsts can be easily rejected on the basis of TOF, while the latter can be excluded by applying a low energy cut in the gamma detector. The choice of applying a 3 MeV energy  cut in our experiment was somehow arbitrary: in principle a 2 MeV cut is enough to exclude the scattered gamma rays and, at the same time, allows preserving PG statistics. However, because of the low energy resolution of our detectors, especially in the case of the BaF$_{2}$, a 3 MeV cut represented a conservative value to guarantee a  more robust analysis. 
The 8 MeV cut applied on our data was also intended to reject neutrons, but it is indeed redundant with a CTR of 100 ps (rms). 
In our experiment, all gamma detectors were of limited size and therefore most of events were detected through Compton interaction in the crystal resulting in a degraded energy response. Nevertheless, the energy resolution was adequate enough to recognise the $\sim$ 3 MeV local minimum in all PG spectra and exclude background events in the low energy region. \\
The method proposed for the PGT spectra analysis is based on the integration of the TOF histograms. This approach has allowed minimising the statistical fluctuations of the curves used to measure the proton range shift, and therefore to increase the sensitivity of the technique with respect to traditional approaches that directly extract different statistical momenta from the PGT spectra. In our approach, an arbitrary threshold is applied to the integral PGT spectra in order to measure the proton range shift. Despite we have verified that small variations of the chosen threshold around the inflection point produces negligible variation of the measured range, we are currently investigating different procedures to make the choiche of this threshold less arbitrary.\\
Finally, we have demonstrated that, in  presence of an heterogeneity, the technique can provide, not only a measure of its thickness, but also a measure of its absolute depth within the target taking the PGT spectra rising edge as reference. This capability has been proven here for an air cavity and an estimated number of $\sim$~3$\times$10$^{8}$ incident protons (from the comparison of simulated and experimental PGT spectra). 
Similar conclusions can be in principle drawn for high density heterogeneities that are detectable as an excess of counts in the PGT spectrum. 

%
\section{Conclusions}
In this paper we have shown, with a combination of experiments and Monte Carlo simulations, that a PGT-based monitoring results in a very good sensitivity in terms of proton range shift thanks to the excellent time resolution achievable in single proton regime. Our results demonstrate that real-time monitoring within a single pencil beam irradiation spot (of 10$^{8}$ protons) is realistic. 
At the same time, the availability of ultra fast (proton and PG) detectors allows losening the energy resolution requirements, paving the way for the use of less complex, smaller and cheaper gamma detectors. 
\section{Acknowledgments}
The authors would like to thank ITMO-Cancer (CLaRyS-UFT project). This work is carried out in the frame of Labex PRIMES (ANR-11-LABX-0063). Part of this work was performed within the framework of the EU Horizon 2020 project RIA-ENSAR2 (654 002)  and is partly supported by the French National Agency for Research called "Investissements d'Avenir", Equipex Arronax-Plus ANR-11-EQPX-0004, Labex IRON ANR-11-LABX-18-01 and ANR-16-IDEX-0007. We are also grateful to the RD42 collaboration members for their useful advice. 
\section*{References}
\begin{harvard}
\item{} Bergonzo P et al., Improving diamond detectors: A device case, Diamond and Related Materials  Volume 16, Issues 4–7, April–July 2007, Pages 1038-1043 
\item{} Breton D et al., Picosecond time measurement using ultra-fast analog memories, Proceedings TWEPP2009, Topical Workshop on Electronics for Particle Physics. Paris 21-25 September 2009
\item{} Gallin-Martel M-L et al., A large area diamond-based beam tagging hodoscope for ion therapy monitoring, EPJ Web of Conferences 170 (09005) (2018)
%
\item{} Golnik G et al. Range assessment in particle therapy based on prompt $\gamma$-ray timing measurements, Phys. Med. Biol.  59(2014) 5399-5422. 
\item{} Hueso-González F et al., First test of the prompt gamma ray timing method with heterogeneous targets at a clinical proton therapy facility, Phys. Med. Biol. 60 (2015) 6247-6272
\item{} Kraan AC, Range verification methods in particle therapy: underlying physics and Monte Carlo modeling, Front. Oncol., 5 (150) (2015) 1-27
\item{} Krimmer J et al., Prompt-gamma monitoring in hadrontherapy: A review, Nuclear Inst. and Methods in Physics Research A 878 (2018) 58-73 
\item{} Kozlovsky B et al., Nuclear Deexcitation Gamma-Ray Lines from Accelerated Particle Interactions, Astrophys. J., Suppl. Ser. 141 (2002) 523–541
%
\item{} Paganetti H, Range uncertainties in proton therapy and the role of Monte-Carlo simulations,  Phys. Med. Biol. 57 (11) (2012) R99-R117 
\item{} Pausch G et al., Detection systems for range monitoring in proton therapy: Needs and challenges, Nuclear Inst. and Methods in Physics Research A, in press 
\item{} Petzoldt J et al., Characterization of the microbunch time structure of proton pencil beams at a clinical treatment facility, Phys. Med. Biol. 61 (2016) 2432-2456 
\item{} Poirier F et al., Studies and upgrades on the C70 Cyclotron Arronax, Proceedings of Cyclotrons 2016, CYC2016, TUDO2, Zurich, Switzerland
\item{} Smeets J et al., Prompt gamma imaging with a slit camera for real-time range control in proton therapy, Phys. Med. Biol. 57 (2012) 3371-3405 
\item{} Vedia V et al., Performance evaluation of novel LaBr3(Ce) scintillator geometries for fast-timing applications, Nuclear Inst. and Methods in Physics Research A 857 (2017) 98-105
\item{} Verburg JM \& Seco J, Proton range verification through prompt gamma-ray spectroscopy, Phys. Med. Biol. 59 (2014) 7089-7106
\item{} Werner T et al., Processing of prompt gamma-ray timing data for proton range measurements at a clinical beam delivery, Phys. Med. Biol. 64 (2019) 105023-105043 
\end{harvard}
\end{document}